\documentclass[twocolumn,showpacs,preprintnumbers]{revtex4} 
\usepackage{amssymb}
\usepackage[dvips]{graphicx}

\begin{document}

\title{Comment  on "Single intrinsic Josephson junction with double-sided fabrication technique" by  You et al [Appl. Phys. Lett. 88, 222501 (2006)]}

\author{V.N. Zavaritsky$^{1,2}$}
\address
{$^{1}$Department of Physics, Loughborough University, Loughborough, United Kingdom, 
$^{2}$Kapitza Physics Institute \& General Physics Institute, Moscow, Russia}


\textbf{Comment  on "Single intrinsic Josephson junction with double-sided fabrication technique" by  You et al [Appl. Phys. Lett. 88, 222501 (2006)]}

In a recent letter, henceforth referred to as Ref.[1], You et al postulate that Bi2212 factually represents a series array of SIS junctions and claim that the nonlinear current-voltage characteristics (IVC) of their bridge are free of heating. Earlier experiments cast serious doubt upon the accuracy of the principal postulate (of this letter), see Ref.[2] and references therein. In what follows I will demonstrate that the major claim by the authors of Ref.[1] is at odds with their own data which suggest an extrinsic cause for the IVC nonlinearities.

The authors of Ref.[1] claim that "Joule heating can easily be ruled out", alleging that $10\mu W$ dissipated in a sample with a twenty-fold difference in area causes the very same heating. This assumption is incorrect as the heat ($W=IV$), dissipated in a sample, escapes through its surface area ($A$). So heating depends on the heat load ($P=W/A$) and, in comparable conditions, $10\mu W$ will cause 20 times higher overheating in a sample of 20 times smaller $A$, see Ref.[2] for details. For this reason, evaluation of heating using $R_{th}$ measured in a sample of different A is not possible unless the area-dependence of the thermal resistance ($R_{th}\propto 1/A$) is taken into account, see Ref.[3] for details. Provided the findings by Ref.[3] are applicable to Ref.[1] (as assumed by its authors), the original $R_{th}$ adjusted for a tenfold area difference between the samples of Refs.[1,3] suggests that $P\simeq 330W/cm^2$ caused by $10 µW$ applied to $A=3\mu m^2$ bridge by Ref.[1] overheats it by 6.5K. A quantitatively similar overheating (8K) is also suggested by the estimates by ref.18 from ref.[1], corrected for a 21-fold $A$-difference. Thus, the authors of Ref.[1] seriously underestimate the heating, which is not at all negligible, since the gap-like feature promoted by Ref.[1] corresponds to the bridge overheating well above $2.5T_B$. 
The estimates above are not extremely accurate as the heat transfer efficiency in Ref.[1] is not the same as in Ref.[3]. Indeed, while the heat escape from the sample of Ref.[3] (and also of Ref.18 in Ref.[1]) was facilitated by a metal electrode of high thermal conductivity, the bridge in Ref.[1] is particularly prone to local overheating as it is sandwiched between a mass of Bi2212, whose poor thermal and electric conductivities are additionally damped by the inevitable strains and cracks introduced into the bulk Bi2212 by its splitting using a "scotch tape" technique by Ref.[1]. For these reasons it is natural to expect that a significantly smaller heat load is required to overheat the bridge by Ref.[1]. 

Ref.[3] provides sufficient experimental means for the quantitative verification of this a-priori conclusion. Indeed, it shows that at sufficiently high heat loads the heating-induced IVC nonlinearities exceed the intrinsic ones so radically that the latter might be safely ignored. The experimental IVC in such circumstances is primarily determined by the normal state resistance, $R_N(T)$, while the mean temperature, $T$,  of the self heated sample is appropriately described by Newton's Law of Cooling (1701),

\begin{equation}
T=T_B+P/h,
\end{equation}
where $T_B$ is the temperature of the coolant medium (liquid or gas) and $h$ is the heat transfer coefficient, which depends neither on $A$ nor $T$, see Refs.[2,4] for details. Ref.[1] presents IVC together with $R(T)$ of the same sample hence allowing a straightforward estimate of the actual heat transfer efficiency. 

\begin{figure}
\begin{center}
\includegraphics[angle=-0,width=0.47\textwidth]{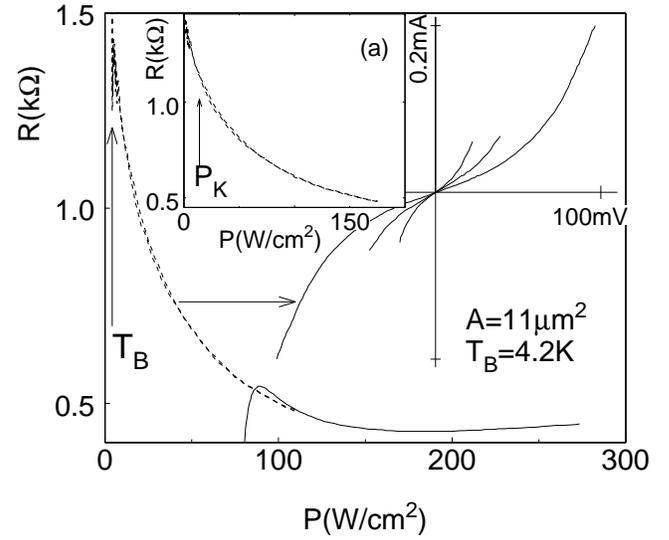}
\vskip -0.5mm
\caption{The solid lines in Fig.1 reproduce the IVC and $R(T)$ of one and the same bridge by Ref.[1]; the broken line in Fig.1b shows the same IVC, re-plotted as a sample resistance, $R=V/I$, versus the mean temperature obtained with Eq.(1) from the heat load, $P=VI/A$; $h\simeq 1.7Wcm^{-2}K^{-1}$. Fig.1a presents the heat load dependence of the bridge's resistance, $R=V/I$, obtained from the same IVC of Fig.3c by Ref.[1]; the dot labelled as $P_K$ marks the typical heat load of a domestic kettle, which is 2-3 orders smaller than P which builds the gap in Ref.[1].}
\end{center}
\end{figure}

As seen in Fig.1, $R_N(T)$ reconstructed from IVC by Ref.[1] using Ohm's law and Eq.(1) correlates reasonably with $R(T)$ of the same bridge, hence suggesting a heating origin for the IVC non linearity and allowing estimation of $h\simeq 1.7Wcm^{-2}K^{-1}$. In remarkable agreement with the a-priori expectations mentioned above, the overall heat transfer efficiency in Ref.[1] is at least an order of magnitude poorer than in Ref.[3] and is about the same as in the very early "mesa" design used by Ref.[5] for example, where heating issues were ignored. Using this $h\simeq 1.7Wcm^{-2}K^{-1}$ we estimate the heating caused by 4-12$\mu W$ dissipated at $T_B=4.2K$ in a $3\mu m^2$ sample as 78-230K, thus confirming the heating origin of the data by Ref.[1].

As far as the $R_N(T)$ is concerned, Fig.1 suggests that $dR_N/dT$ remains negative throughout the temperature range below $T^*$, the temperature at which the out-of-plane resistance reaches its minimum. Such behaviour agrees reasonably with the direct $R_N(T)$ measurements under conditions where the superconductivity of Bi2212 was suppressed by a high magnetic field (see Ref.[6] and references therein) or current  Ref.[7]. 

It can be seen, therefore, that the conclusions by the authors of Ref.[1], like the similar findings discussed in Refs.[2-4],  are not beyond dispute as their experimental results are most likely caused by self-heating. 

V.N. Zavaritsky

Department of Physics, Loughborough University, UK; Kapitza Physics Institute and General Physics Institute,  Moscow


\end{document}